\documentstyle[epsf,aaspp4, 12pt]{article}

\lefthead{Scaling Relations for Gravitational Clustering in Two Dimensions}

\righthead{J.S.Bagla, S.Engineer and T.Padmanabhan}

\begin{document}

\title{Scaling Relations for Gravitational Clustering in Two Dimensions}
\author{J.S.Bagla$^1$, S.Engineer$^2$ and T.Padmanabhan$^2$}

\affil{$^1$Institute of Astronomy, University of Cambridge,\\
Madingley Road, Cambridge CB3 0HA, U.K.\\
$^2$Inter-University Centre for Astronomy and Astrophysics,\\
 Post Bag 4, Ganeshkhind, Pune 411 007, INDIA}

\authoremail{jasjeet@ast.cam.ac.uk, sunu@iucaa.ernet.in and 
paddy@iucaa.ernet.in}

\begin{abstract}
It is known that radial collapse around density peaks can
explain the key features of evolution of correlation function in
gravitational clustering in {\it three} dimensions. The same model
also makes specific predictions for {\it two} dimensions.   In this
paper we test these predictions  in two dimensions with the help of N-Body
simulations.  We find that there is no stable clustering in the
extremely non-linear regime, but a nonlinear scaling relation does
exist and can be used to relate the linear and the non-linear
correlation function. In the intermediate regime, the simulations
agree with the model. 
\end{abstract}


\section{Gravitational Clustering: Two vs Three Dimensions}

Evolution of density perturbations at scales smaller than the Hubble
radius in an expanding Universe can be studied in the Newtonian limit
in the matter dominated regime.  Linear theory is used to study the
growth of small perturbations in density but a study of non-linear
clustering requires N-Body simulations.  A number of attempts have
been made in recent years to understand the evolution of constructs
like the two point correlation function using certain non-linear
scaling relations (NSR). [See for example (\cite{hamilref});
(\cite{rntp}); (\cite{tpgdeu}).]  These studies have shown that the
relation between the non-linear and the linearly
extrapolated correlation functions is reasonably model independent.
This relation divides the evolution of correlation function into three
parts (\cite{apmpap}): the linear regime, the intermediate regime and the 
non-linear regime.  The evolution in the intermediate regime can be
understood in terms of radial collapse around density peaks
(\cite{tpgdeu}), if it is assumed that the evolution of profiles of
density peaks follows the same pattern as an isolated peak.  It is
customary to invoke the hypothesis of stable clustering (\cite{lssu})
to model the non-linear regime.  A large number of studies have
examined clustering in this regime and the general consensus is
that the stable clustering limit does not exist (\cite{tpcen}).

However, the limited dynamic range of currently available
3-dimensional N-Body simulations poses serious difficulties in
investigating this problem in greater detail.  It was pointed out
(\cite{tpdonald}) that we can circumvent this problem by simulating a
two dimensional system, wherein a much higher dynamic range can be
achieved. For example, since $160^3\approx 2048^2$, the computational
requirements are the same for a 2D simulation with box size of 2048
and 3D simulation with box size of 160. Assuming that one can reliably
use, say, half of box size as {\it good} dynamic range we have a
dynamic range of factor 1000 in 2D against a factor of about 80 in
3D. This allows us to probe higher nonlinearities in 2D compared to
3D. As long as we stick to generic features (like the non-linear
scaling relations, investigated here) which are independent of
dimension, 2D has a definite advantage over 3D.  Higher dynamic range
is the basic motivation for studying gravitational clustering in two
dimensions. 

When we go from three to two dimensions, we have, in principle, two
different ways of modeling the system:
\par
(i) We can consider two dimensional perturbations in a three
dimensional expanding Universe. Here we keep the force between
particles to be $1/r^2$ and assume that all the particles, and their
velocities,  are confined to a single plane at the initial
instant. 

(ii) We can study perturbations that do not depend on  one of the
three coordinates, i.e., we start with a set of infinitely
long straight ``needles'' all pointing along one axis. The
force of interaction falls as $1/r$. The evolution keeps the
``needles'' pointed in the same direction and we study the clustering
in an orthogonal plane.  Particles in the N-Body simulation represent
the intersection of these ``needles'' with this plane. 
In both these approaches the universe is three dimensional and the
background is expanding isotropically. 

The study of 2-D perturbations (like those due to pancakes, for example)
in a 3-D expanding Universe faces an operational problem: To begin
with, we do not gain the dynamic range if we stick to 3D, even if we
consider perturbations in a plane; the force between particles still
has to computed by the solution of Poisson equation in three dimensions. 
Also, relevance of the interaction of matter outside the plane
with these perturbations makes it, essentially, a 3D problem.

Thus we are left with the second possibility. The two dimensional
system is the intersection of an orthogonal plane and the ``needles''
and the force between the  ``particles''  in this plane is given by
the solution of the Poisson equation in two dimensions.  Such 
a system is somewhat dichotomous with the background universe
expanding isotropically.  However, convenience is not the only reason
for studying this somewhat strange system --- relevant results for the
evolution of density profiles around peaks in 2-D have also been
computed for this type of a system (\cite{fg_selfsim}).  

Generalization of the NSR to the 2-D system was done using
relations for cylindrical collapse by Padmanabhan (1996b) and we will
test these predictions here.

Although the system of infinite needles is appropriate for testing the
predictions in the intermediate regime, the same cannot be said for
the asymptotic regime.  We are dealing with a system that occupies a
smaller number of dimensions in the phase space {\it and} the
interaction of the  constituents follows a different force law.
Therefore, it is difficult to interpret, or carry over, results
regarding stable clustering to the full 3-D system. 

\subsection{Non-linear Scaling Relations}

The non-linear and the linear correlation functions at two different
scales can be related by NSR.  The relation between these scales is
given by the characteristics of the pair conservation equation
(\cite{rntp}).  For the two dimensional system of interest, this
equation can be written as (\cite{tpdonald}) 
\begin{equation}
{\partial D \over \partial A} -h (A, x) {\partial D \over \partial X}
= 2h (A, X), 
\end{equation}
Here $D=\log(1 + \bar\xi)$, $h = - v_p/Hr$ is the scaled pair
velocity, $\bar\xi(x)=2 x^{-2} \int^x r \xi(r) dr$ is the mean
correlation function ($\xi$ is the 
correlation function),  $H$ is the Hubble's constant, $X=\log(x)$ and
$A=\log(a)$.  The characteristics
of this equation are $x^2 (1 + \bar \xi(x,a)) = l^2 $, where $x$ and
$l$ are the two scales used in NSR.  The self similar models due to
Filmore and Goldreich (1984) imply that for collapse of cylindrical
perturbations the turn around radius and the initial density contrast
inside that shell are related as $x_{\rm ta} \propto
l/\bar\delta_i \propto l/\bar\xi_L (l)$.  (Here $\bar\xi_L$ is the
linearly extrapolated mean correlation function).  Noting that in
two dimensions $M \propto x^2$, we find $\bar\xi (x) \propto
\left[ \bar\xi_L (l) \right]^2$ in the regime dominated by infall. 
Stable clustering limit implies $\bar\xi_{NL}(a, x)
\propto \bar\xi_L(a, l)$ (\cite{tpdonald}).  Thus in 2-D the scaling
relations are
\begin{equation}
\bar \xi (a,x) \propto \left\{ \hbox{  }
\begin{array}{ll} 
\bar \xi_L (a,l) & \hbox{ (Linear)} \\
\bar \xi_L(a,l)^2  & \hbox{ (Radial Infall)} \\
\bar \xi_L(a,l) &  \hbox{ (Stable Clustering)} \\
\end{array} \right.
\label{hamilton}
\end{equation}
A more general assumption compared to stable clustering involves
taking $h=$~constant asymptotically. In a system reaching steady state with
both virialisation and mergers contributing to the evolution, one may reach
a constant value for $h$, though it will not be unity if mergers are a
dominant phenomenon. (This assumption has been discussed in, for example, 
\cite{tpgdeu},1996b,1997.)
It also allows a larger parameter space to compare simulation results.
If $h$=constant asymptotically, then $\bar\xi(x) \propto {\bar\xi_L}^h(l)$ 
in this limit.  Note that in 3D, the indices for three regimes are 
$1$, $3$ and $3h/2$ respectively.

All features of clustering in three dimensions are present
here as well. In particular,
\par
(i) If the asymptotic value of $h$ scales with $n$ such that $h (n +
2 )= {\rm constant}$ then the final slope of the non-linear correlation
function will be independent of the initial slope.
\par
(ii) If NSR exists then it will predict a {\it specific} index in the
intermediate and asymptotic regimes which will depend on the initial
power spectrum.  In other words, existence of NSR implies that
gravitational clustering does not erase memory of initial conditions.
\par
(iii) It is, however, possible that spectra which are not scale free
acquire universal critical indices at which the correlation
functions grow in a `shape invariant' manner. This comes about because
the growth rate of correlation function varies with the local index
and for an index that is not globally constant the correlation functions
may `straighten out' by this process.
\par
(iv) In 3-D clustering, $n=-1$ in the intermediate regime and $n=-2$
in the asymptotic regime (\cite{critindex}) are the critical indices.
These are the same for clustering in two dimensions. 

\section{Simulations and Results}

We carried out a series of numerical experiments to test the ideas
outlined above.  We used a particle mesh code (\cite{pmcode}) to
simulate power law models.  The simulations were done with $1024^2$ or
$2048^2$ particles in order to ensure that we had sufficient dynamic
range to study all the three regimes in evolution of non-linear
clustering.  In particular, it is necessary to use larger simulations
for power law spectra with a negative index.  Here, we will present
results for three models: $n=1$, $n=0$ and $n=-0.4$. 

All the models are normalized by requiring the linearly extrapolated
root mean square fluctuations in density, computed using a Gaussian
filter, to be unity at a scale of $10$ grid points at $a=1.0$.
The results we present are for $a=1$, $2$ and $5$ for $n=0$ and $n=1$,
and $a=1$, $2$ and $3$ for $n=-0.4$.

A significant source of errors in large simulations is the addition of
a small displacement in each step (fraction of a grid length) to a
large position (up to 2048 grid lengths).  We avoid this problem by
using net displacement for internal storage.  

We will show the correlation function and the pair velocity only for
length scales larger than four grid lengths.  We do this to avoid
error due to shot noise and other artifacts introduced by various
effects at smaller scales.  This ensures that errors in our results
are acceptably small.  (Variations between different realizations give
a dispersion of less than $10\%$ in the correlation function.)

In fig.1 we have plotted the non-linear correlation function
$\bar\xi(x)$ as a function of the linearly extrapolated correlation
function $\bar\xi_L(l)$.  Here the scales $x$ and $l$ are related by
$x^2 (1 + \bar \xi) = l^2 $.  Data for $n=1$ is represented by
circles, that for $n=0$ by stars and `$+$' marks the points for
$n=-0.4$.  Clearly, there are no systematic differences between
the three models and the data points trace out a simple curve with
three distinct slopes (We have also marked the 2 $\sigma$ errors
calculated by averaging over several data sets. The error bars  are plotted 
away from the NSR plot, for visibility and clarity.). The NSR, shown as thick lines, is
\begin{equation}
\bar \xi (a,x) = \left\{ \hbox{  }
\begin{array}{ll} 
\bar \xi_L (a,l) & \hbox{$\bar\xi_L(l) \leq 0.5$; $\bar\xi(x)\leq 0.5$}\\
 2 {\bar\xi_L(a,l)}^2  & \hbox{$0.5 \leq \bar\xi_L(l) \leq 2$; $0.5 \leq
\bar\xi(x)\leq 8 $} \\
 4.7 {\bar\xi_L(a,l)}^{3/4} &  \hbox{$2 \leq \bar\xi_L(l)$; $8 \leq 
\bar\xi(x)$ } \\
\end{array} \right.
\label{ourfit}
\end{equation}
The slope in the intermediate regime is as expected.  The asymptotic
regime has a different slope than that predicted by stable clustering,
which is shown as a dashed line.
Unlike the observed relations for clustering in three dimensions, the
coefficient for the intermediate regime is large.  This has
important implications for the critical index.  

Panels of fig.2 show $\bar\xi(x)$ as a function of $x/x_{nl}$ for
the three models.  These confirm that the slope of $\bar\xi(x)$
is consistent with the NSR shown in fig.1.  In each of these
panels, the slope expected in the stable clustering limit is shown as
a dashed line.

As mentioned above, the existence of the NSR (eqn.(\ref{ourfit}))
implies that the slope of the correlation function will depend on the
initial  spectral index. To this extent, gravitational clustering does
not erase memory of initial conditions.  However, the differences of
slope are significantly reduced by non-linear evolution.

\section{Conclusions}

Our conclusions can be summarized as follows:
\par
(i) We have verified  that NSR for the correlation function exist
for clustering in 2D in all the three regimes, just like in 3D. This
NSR  is independent of the power law index -- at least for the three
indices studied here.
\par
(ii) In the intermediate regime, the NSR in the form of
eqn.(\ref{ourfit}), can be understood in terms of radial infall
around peaks. Our simulations verify the predictions (\cite{tpdonald})
for this regime.  
\par
(iii) In the asymptotic regime, our results do {\it not} agree with
the stable clustering hypothesis.  The slope of the NSR in the
asymptotic regime in fig.1 implies $h=$ constant.  We find that, in
this regime, $h \simeq 3/4$ for all the models studied here.
\par
(iv) The existence of NSR implies that the asymptotic slope of the
correlation function depends on the initial slope.  However, this is
strictly true only for pure power law models; for other models it is
possible for the spectra to be driven to a universal form.  

The NSR in the asymptotic regime seems to be linked to the logarithmic
nature of the potential.  Issues relating to theoretical modeling of
this regime will be addressed in a future publication. 


While this paper was in preparation, a preprint (\cite{dipak}) that
discusses similar issues appeared on SISSA archives.  However our
results are  different from theirs in several aspects: (a)
we find a model independent NSR with an asymptotic slope of $3/4$
whereas Munshi et al (1997) only report deviations from stable
clustering. (b) We do {\it not} find that $h(n+2)$=constant
is a good fit to our data. They seem to conclude differently even though
their figure 2 shows a large scatter. Their fit to
$h(n+2)$=constant is also not good and the omission of the first point
will make their fit consistent with a constant asymptotic value for $h$
around $0.5-0.75$. (c)  Lastly, a comparison of our figure
1 with the top panel of figure 1 in their paper shows that whereas we
get the same transition points between the three regimes for all the
models, the transition points deduced by them tend to vary between models.
The differences can possibly be understood
as arising from lower resolution and inadequate levels of
non-linearity in their simulations.

\section{Acknowledgments}

We thank the anonymous referee for useful comments.  JSB acknowledges
the support of the PPARC fellowship at the Institute of Astronomy,
Cambridge, U.K. SE thanks SERC for support during the course of this
work.

\begin{figure}
\epsfxsize=5truein\epsfbox[40 28 502 544]{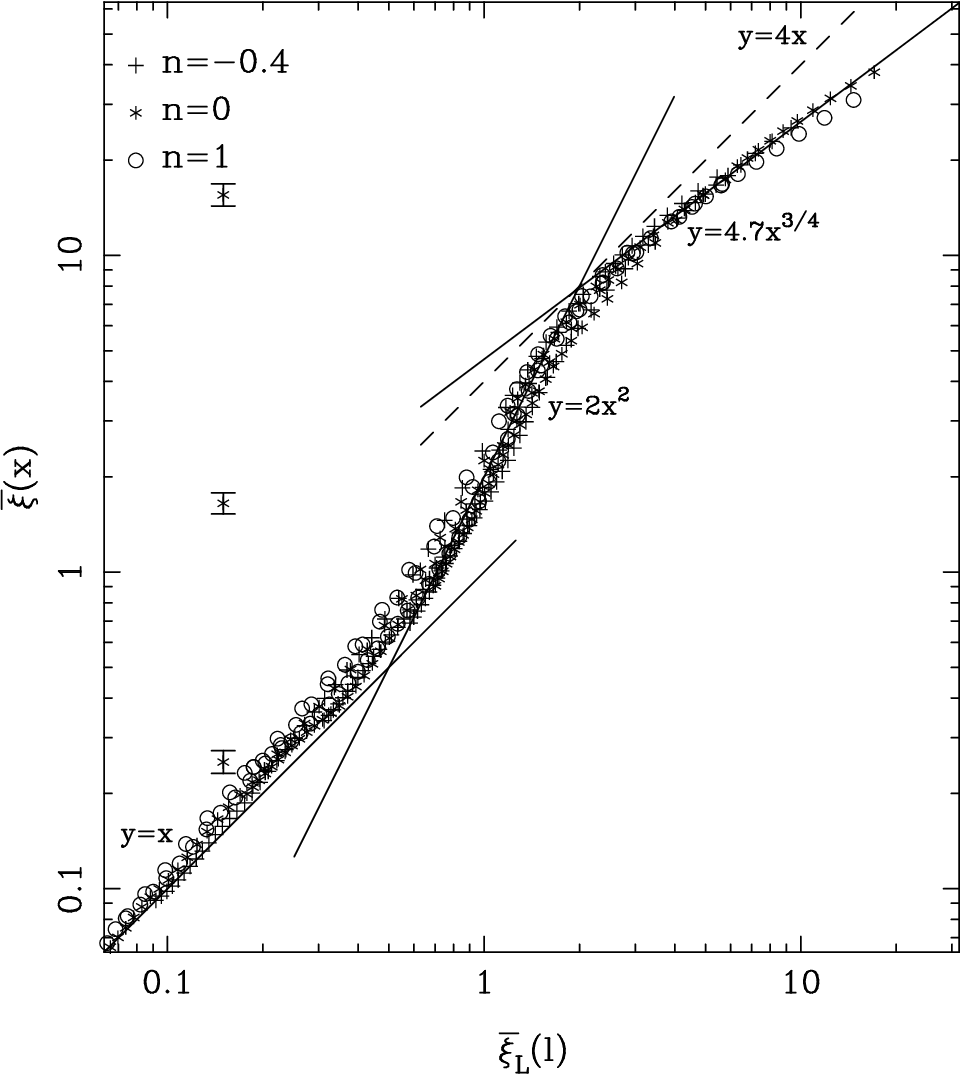}
\caption{This figure shows the non-linear correlation function
$\bar\xi(x)$ as a function of the linearly extrapolated correlation
function $\bar\xi_L(l)$.  Here the scales $x$ and $l$ are related by
$x^2 (1 + \bar \xi) = l^2 $.  Data for $n=1$ is represented by
circles, that for 
$n=0$ by stars and $+$ marks the points for $n=-0.4$.  For each of
these models we have plotted data for the three epochs mentioned in
the text.  The estimated 2 $\sigma$ error bars are shown as vertical lines at  three representative values of $\bar\xi$ {\it viz.} at $\bar\xi$=15.582, 1.65 and  0.25, covering the nonlinear, intermediate and linear regimes. The error bars are shown away from the NSR plot for the sake of visibility. It is clear
from this figure that there are no systematic differences between
the three models and they trace out a simple curve with three distinct
slopes.  The slope of the curve in the intermediate regime is same as
that predicted by the radial infall model.  The stable clustering
limit is shown as the dashed line and it is clear that the data points
deviate from this curve.}
\end{figure}

\begin{figure}
\epsfxsize=5.5truein\epsfbox[41 24 502 562]{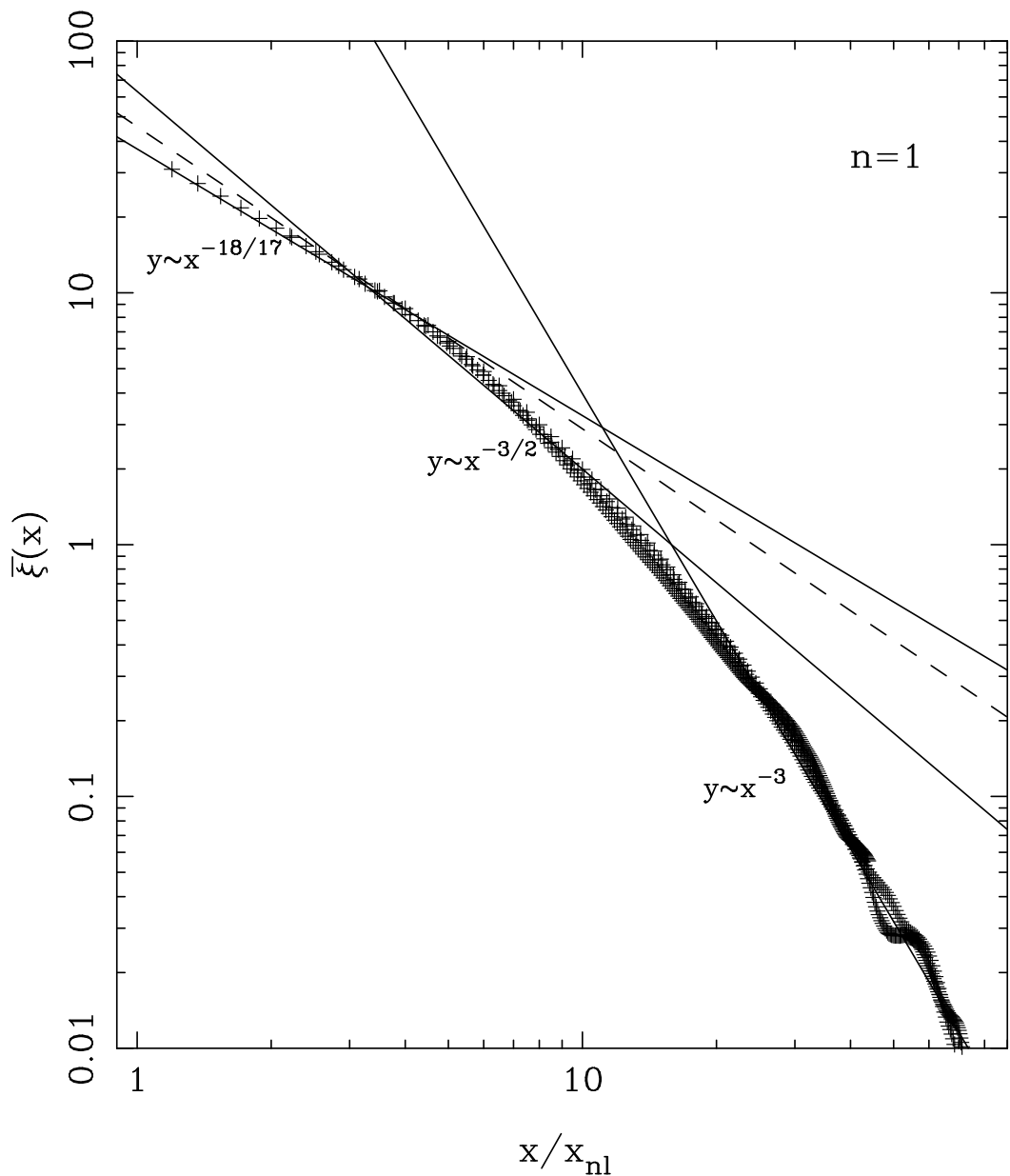}
\caption{This figure shows the correlation function $\bar\xi(x)$ as a
function of $x/x_{nl}$ for $n=1$ model.  Here $x_{nl} \propto
a^{-2/(n+2)}$.  Thick lines mark slopes
expected from the non-linear scaling relations shown in figure 1.  The
dashed line marks the expected slope of the correlation function in
the stable clustering limit.  The mismatch between the expected slope
and the true slope in the intermediate regime may arise from the fact
that the assumption of $\bar\xi \gg 1$ used in computing the slope is
not valid at the lower end of the regime.}
\end{figure}

\setcounter{figure}{1}

\begin{figure}
\epsfxsize=5.5truein\epsfbox[40 24 502 562]{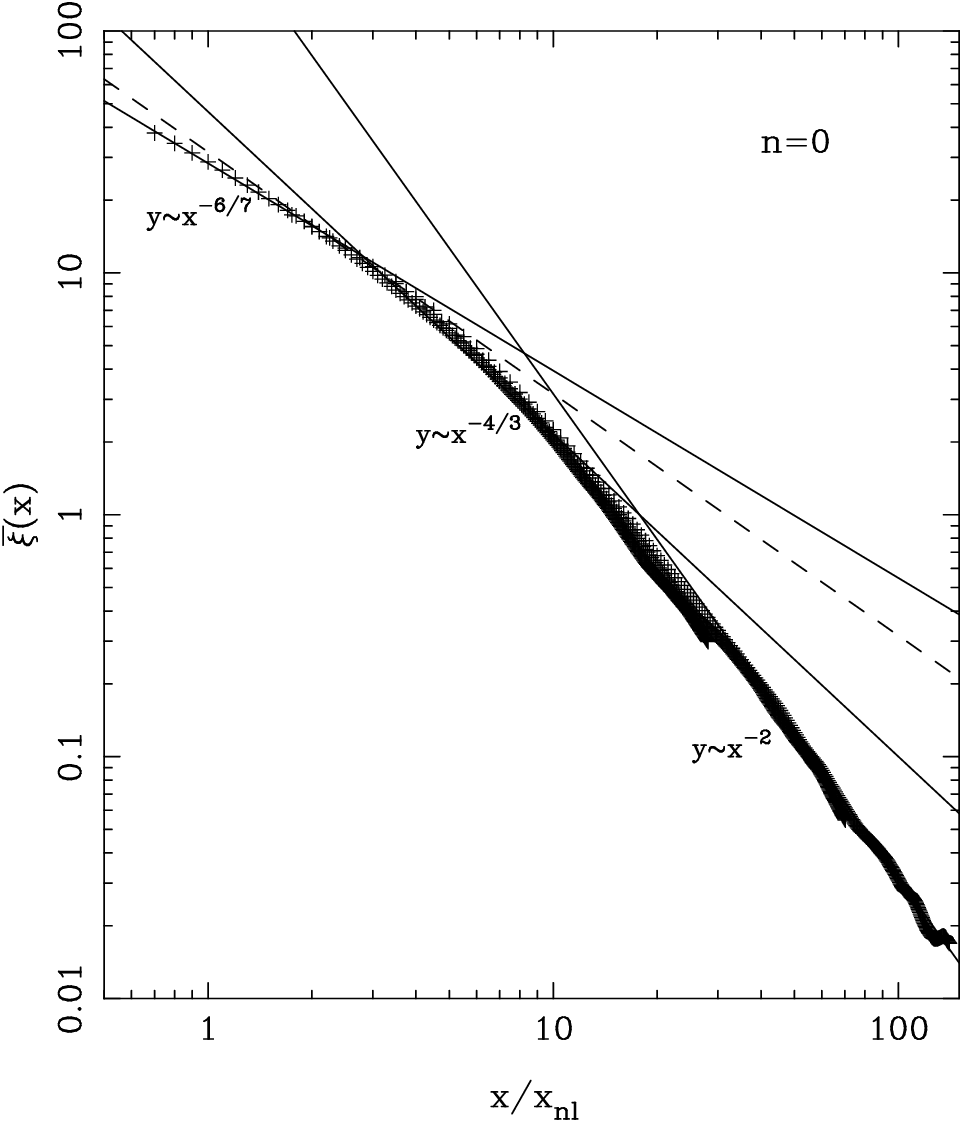}
\caption{Continued.  This panel shows the same plot for $n=0$.}
\end{figure}
\setcounter{figure}{1}

\begin{figure}
\epsfxsize=5.5truein\epsfbox[40 24 502 548]{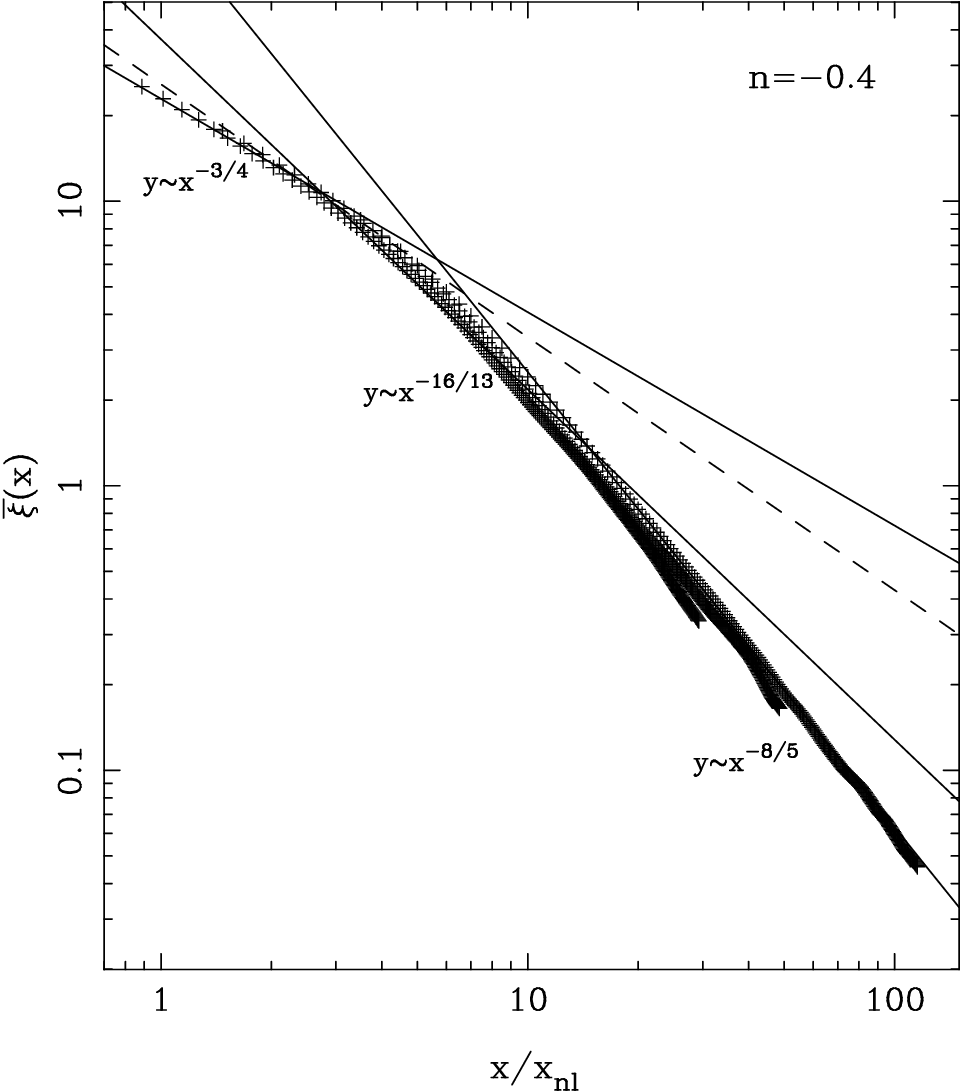}
\caption{Continued.  This panel shows the same plot for $n=-0.4$.}
\end{figure}

\end{document}